# Machine Learning-Based Evaluation of Attitude Sensor Characteristics Using Microsatellite Flight Data

By Yuji SAKAMOTO[1)]

[1)]*Green Goals Initiative, Tohoku University, Sendai, Japan*

Using actual flight data from a 50-cm-class microsatellite whose mission and operations have already been completed, this study re-evaluates satellite attitude determination performance and the error characteristics of onboard attitude sensors. While conventional approaches rely on batch estimation or Kalman filtering based on predefined physical models and white-noise assumptions, this research introduces a machine-learning-based approach to extract and correct structural and nonlinear error patterns embedded in real observational data.

In this study, high-quality attitude determination results obtained from star sensors and a fiber optical gyro (FOG) are treated as ground truth, and machine learning is applied to coarse attitude sensor data consisting of Sun sensors and magnetic field sensors. A one-dimensional convolutional neural network (Conv1D) is employed to regressively predict attitude from short sequences of time-series sensor measurements. The model is trained and evaluated using five sets of on-orbit observation logs, with four passes used for training and one independent pass used for testing. The results show that, while conventional coarse attitude determination using the TRIAD method yields attitude errors on the order of 7 deg RMS, the proposed machine-learning approach achieves RMS errors of approximately 0.7 deg for the training data and 2–3 deg for the test data, depending on the sensor combination.

**Key Words:** attitude determination, machine learning, convolutional neural network, coarse attitude sensors, microsatellites

## 1. Introduction

Tohoku University has accumulated extensive experience in the development and operation of multiple 50-cm-class microsatellites, beginning with the SPRITE-SAT/RISING satellite, which has been in operation since 2009. Many of these satellites are equipped with optical telescopes and multi-wavelength sensors for Earth observation and are operated by continuously acquiring images while maintaining attitude tracking toward a specific ground location designated by command.

In the most recent satellite, RISING-4 (development name; Fig. 1), attitude determination was primarily performed using star sensors and a fiber optical gyro (FOG). Because the star sensors were in-house developments, their design specifications caused attitude detection to cease even when a small amount of light from the Sun, Moon, or Earth entered the sensors. As a result, attitude angles detected over a short duration were frequently used as initial values, followed by extensive reliance on attitude integration using the FOG. An example of attitude analysis for this satellite is presented in Ref. [1].

In addition, the same type of Sun sensors and magnetic field sensors has been continuously used since the first satellite. Although attitude determination using the TRIAD method involves relatively large errors, it has been effectively utilized for purposes such as guiding the spacecraft toward an initial attitude suitable for star sensor acquisition.

Because mission requirements had been successfully achieved using classical filtering approaches based on predefined motion models, there had been no opportunity to replace them with AI technologies, including machine learning (ML), which consume additional computational resources. However, if it is possible to achieve high-accuracy predictions based directly on actual observations without explicitly defining models—particularly for components that were previously simplified as random errors due to unknown or unmodeled effects—such approaches should be actively adopted.

Furthermore, significant benefits are expected in terms of reducing human workload and operational time during satellite development and operations. Proposals for satellite systems assuming cooperative operation among multiple satellites are becoming increasingly active, creating a need for efficient simultaneous operation technologies. To reduce launch costs and prevent space debris, deployment opportunities at lower altitudes are becoming common, raising concerns that Sun-synchronous orbits may deteriorate within approximately three years. Under such conditions, initial operations must be completed rapidly with limited personnel and time. In these situations, solutions based on AI technologies become indispensable.

In this study, machine learning techniques that were not envisioned during the development or operation phases are applied to observation data obtained from a 50-cm-class microsatellite whose mission has already been completed. The advantages of this approach and its potential future applications are discussed. The satellite is equipped with star sensors and a FOG, providing high-quality attitude determination data with relatively small errors. These data are interpreted as ground truth, and the measurement errors and alignment characteristics of the Sun sensors and magnetic sensors are re-evaluated. The insights obtained through this analysis are expected to contribute to the development and utilization of onboard AI computing systems for future satellites.

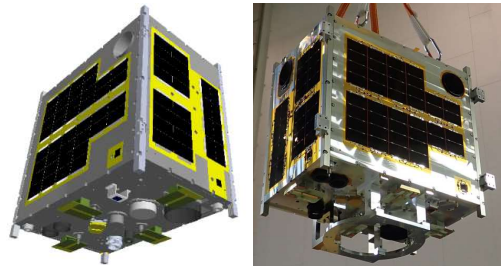

Fig. 1. Appearance of the microsatellite

In this ML analysis, a Conv1D (one-dimensional convolutional neural network) model is employed, in which time-series sensor data at consecutive time steps are used as inputs. This configuration enables the extraction of features that reflect local temporal variations and allows attitude values to be predicted in a regression framework (Ref. [2]).

## 2. Overview of Satellite Data and Machine Learning Model

### 2.1. Satellite characteristics

The RISING-4 satellite was operated for approximately two years, from 2021 to 2023, in an orbit following deployment from the International Space Station. The onboard attitude and navigation sensors include star sensors (developed by Tohoku University, quaternion output), a fiber optical gyro (FOG; manufactured by Tamagawa Seiki CO., LTD), Sun sensors (voltage measurements obtained from fixed resistors connected to 2 × 2 cm power-generation solar cells attached to six external panels), and magnetic field sensors (±2 gauss range, voltage output). In addition, the satellite is equipped with a GPS receiver, four reaction wheels (manufactured by Tamagawa Seiki), and three-axis magnetic torquers (Fig. 2).

The power generation capability of the satellite is 43.9 W during spin operation, and no deployable solar panels are installed. Under nominal conditions, the satellite is maintained in a low-power mode consuming 16.9 W. Continuous observation is not performed; instead, imaging observations are conducted only when the satellite passes over designated regions during daylight (power consumption: 55.9 W), and high-speed communications are performed only during nighttime passes over ground stations (power consumption: 54.4 W, 20 Mbps), which constitute the standard operational profile.

The primary mission sensors are a high precision telescope (HPT; 2.2 m ground resolution, 3.6 × 2.7 km per image) and a space-borne multispectral imager (SMI with LCTF; 47 m ground resolution, 77 × 58 km per image). The SMI is equipped with two LCTFs (430–740 nm and 730–1020 nm), enabling the selection of a total of 590 wavelength bands at 1 nm intervals. Because one wavelength band is selected per image, continuous imaging is performed while switching wavelengths. Consequently, the satellite is required to maintain pointing toward the target location for a certain duration. In addition, a medium field camera (MFC; 35 m ground resolution, 58 × 44 km field of view) and a wide field camera (WFC; 7 km ground resolution, 180 × 134 deg field of view) are installed to support attitude determination.

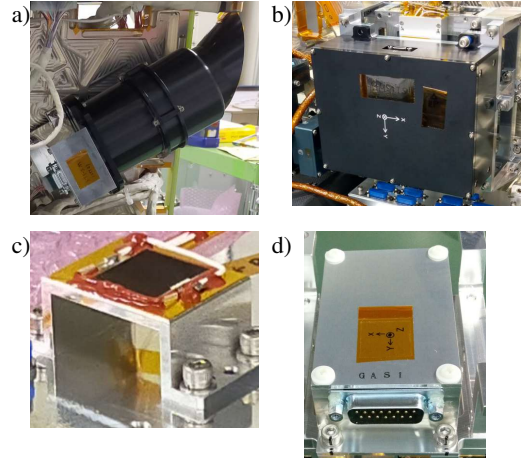

Fig. 2. Appearance of the attitude sensors (a: star sensor, b: fiber optical gyro, c: Sun sensor, d: magnetic field sensor)

During a single sunlit period (approximately 60 minutes), only one imaging target is observed. At the beginning of the sunlit period, the influence of sunlight and Earth-reflected light is relatively small, resulting in a stable probability of star sensor detection. Using coarse attitude determination based on Sun sensor and magnetic sensor measurements (TRIAD method), the satellite is controlled to orient toward a direction where the density of visible stars is high and optical disturbances from the Sun, Moon, and Earth are minimal at that time. Within a predefined time window, the star sensors achieve attitude acquisition, after which a large attitude maneuver is executed so that the mission sensors point toward the designated ground target. During this process, attitude is also computed in parallel by integrating the angular rate output from the FOG; therefore, the star sensor output is allowed to stop. In the majority of operations, the FOG-integrated attitude is prioritized. In extreme cases, the imaging operation can be successfully completed even if the star sensors detect attitude at only a single point in time (1 s).

The attitude computer is capable of logging data with specified sampling intervals and total durations. As a standard procedure, data are recorded for 6 minutes (1 s interval, approximately 360 data points) for each imaging pass, from star acquisition to the completion of imaging. In addition to measured values (GPS, coarse attitude sensors, and fine attitude sensors), onboard estimation results (attitude determination for each mode) and reaction wheel control histories are also stored. This enables evaluation of the differences between newly proposed attitude determination methods using raw sensor data and the attitude determination actually performed onboard.

### 2.2. Data set preparation

A total of five sets of log data obtained during imaging observations over Japan are used in this study. All data correspond to periods encompassing an attitude maneuver followed by telescope imaging.

Each pass consists of 362 data points sampled at 1 s intervals. The attitude sensor measurements and model-derived quantities calculated from satellite position and

time—such as the Sun direction, magnetic field direction, and Earth direction—are defined as the data set X. The attitude determination results are defined as the data set Y.

The attitude determination values recorded in the log data are provided in quaternion form with four elements at each time step. These are converted into Modified Rodrigues Parameters (MRP) and reduced to three elements. This conversion is performed for compatibility with other analysis tools; the advantages or disadvantages of this representation for machine learning have not been evaluated.

The first four passes are used for training (X_train, Y_train), and the remaining one pass is used for testing (X_test, Y_test). Cross-validation is not performed.

The onboard attitude values obtained through star sensor measurements and FOG integration are treated as ground truth (Y_train, Y_test). Using the trained ML model, attitude values (Y_pred) are estimated from the test measurement data (X_test), and the error is evaluated by comparing Y_pred with the ground truth Y_test using the root-mean-square (RMS) value.

Sun sensor measurements are denoted as CSS (sensor-fixed frame c; sensors installed on each external panel, yielding a total of six elements at each time step), and magnetic sensor measurements are denoted as MAG (sensor-fixed frame m, three elements at each time step). Unit vectors of the Sun direction and magnetic field direction in the Earth-centered inertial (ECI) frame, calculated from time and orbital position (obtained by GPS) using mathematical models, are denoted as uS_i and uB_i, respectively. The quantities CSS, MAG, uS_i, and uB_i are included in the operational attitude log data.

Angular rate measurements W (sensor-fixed frame g) are referenced from the operational log data. The vector W_g, obtained by scaling all elements over all time steps so that the maximum value becomes 1, is used.

Both CSS and MAG are positive integer values obtained by ADC voltage measurements. For MAG, values larger than a predefined reference indicate a positive magnetic field direction, whereas smaller values indicate a negative direction (in gauss). This reference value can be determined from preflight ground measurements and operational data. The converted unit vector is defined as uB_m (three elements). For CSS, the larger value is extracted from each of the ±X, ±Y, and ±Z components in the sensor-fixed frame c and converted into a unit vector, defined as uS_c (three elements). Conversely, the smaller value is extracted and defined as uE_c (three elements), where S denotes Sun and E denotes Earth. Although all original values are positive, the sign is inverted when extracting values corresponding to the negative axis. This processing is based on the assumption that larger values represent direct sunlight, whereas smaller values represent Earth-reflected albedo light. In conventional analyses, components corresponding to uE_c were omitted; however, in the ML training, they are included in X with the expectation that they may contribute to attitude determination.

In the present ML analysis, no recalculation using a satellite simulator is performed. Instead, CSS (converted to uS_c and uE_c), MAG (converted to uB_m), uS_i, uB_i, uE_i (= - r_i), and W_g are included in the data set X directly from the operational log data. The maximum dimensionality of X is 21 elements (= 3 × 7).

Some vectors included in X form strongly related pairs (e.g., uS_c and uS_i); however, their physical relationships are not explicitly provided to the ML model. Instead, it is expected that the ML algorithm will automatically learn these patterns.

### 2.3. Overview of the machine learning method

Conv1D is a method that applies convolutional operations to numerical time-series data in order to extract characteristic local patterns within the sequence.

Each pass (four passes in total for training) contains continuous data at time steps t[0] .. t[361]. Because all data strictly maintain a 1 s sampling interval, time itself is not provided to the ML model. Let the raw data sequences be denoted as x_r[0] .. x_r[361] and y_r[0] .. y_r[361] (r = raw). For each y_r, the model is trained to learn its relationship with the preceding n time steps of x_r. When n = 11, the first training set is defined as X[0] = x_r[0] .. x_r[10] and Y[0] = y_r[10]. The next set is X[1] = x_r[1] .. x_r[11] and Y[1] = y_r[11]. From the 362 time steps in each pass, a total of 352 training sets are generated. Because sequential training in time order may introduce bias into the model, all sets are randomly shuffled before training. In the analyses described later, the temporal window length ( n ) is adjustable within the range ( n ≦ 11 ); based on performance comparisons, n = 5 is adopted as the baseline configuration.

The loss function is defined as the root-mean-square (RMS) value of the rotation angle between Y_true and Y_pred for the attitude parameter mrp (three elements) over the output sequence Y[0] .. Y[351].

The maximum number of training epochs is set to 240, and training is terminated when the average loss over the most recent 40 epochs increases. The model corresponding to the minimum loss is selected as the final model. If matrix divergence occurs during training, the model parameters are reverted to those from two steps earlier, the learning rate is multiplied by 0.9, and the optimizer is reinitialized before resuming training.

Through Conv1D processing, a nonlinear model is constructed to transform X (21 elements with up to 11 time steps) into Y (3 elements). For comparison, in a linear transformation model that expands 21 elements to 64 features, the parameters H01 and b0 in ( X1 = H01 X0 + b0 ) would be identified. In the present model, the feature dimensions are expanded from X0(21) to X1(64) and then to X2(128), followed by compression from X2(128) to X3(64) and finally to Y(3) to obtain the output Y_pred. By inserting ReLU functions (values ≥ 0 remain unchanged, values < 0 are set to 0) in each "->" process, the mapping from X_0 to Y is modeled as a complex nonlinear function. The weight matrices H01, H12, H23, H3y and biases b0, b1, b2, b3 are optimized so as to minimize the RMS loss. To prevent overfitting and stabilize training, a dropout rate of 1% is applied only between the final layers X3(64) and Y(3).

## 3. Analysis Results

### 3.1. Evaluation data

Log data obtained from a total of five imaging observations conducted over Japan during a seven-day period in December 2021 are used. To facilitate a preliminary evaluation of the characteristics of the ML analysis, observation logs from the same or similar regions are intentionally used for training, providing favorable conditions for identifying similarities in the data.

The overview of each pass (time and observation target) is as follows:
P1: 2021/12/18, 4 h UTC, Sendai
P2: 2021/12/21, 3 h UTC, Nagoya
P3: 2021/12/22, 2 h UTC, Sendai
P4: 2021/12/23, 3 h UTC, Shimonoseki
P5: 2021/12/24, 2 h UTC, Aoga-Shima

Because the orbit follows deployment from the International Space Station, the local solar time is not fixed. However, the sample passes occur between approximately 02:00 and 04:00 UTC (11:00–13:00 JST), during which the relative geometry among the satellite, the Sun, and the Earth is relatively stable. In addition, because these orbits pass near local noon, the Sun and Earth directions relative to the satellite are nearly opposite, allowing the Sun sensors (six directions in total) to separately measure direct sunlight and Earth-reflected albedo light.

Figure 3 shows the time histories of the maneuver angles to the target for each pass, derived from onboard data. Because the target attitude is selected to facilitate star sensor acquisition and data recording begins after that condition is achieved, the initial values are close to 0 deg. At approximately 1 min after the start of the record, the target is changed and an attitude maneuver begins; in all cases, the maneuver is completed by approximately 3 min. Since the same attitude maneuver strategy is applied to all passes, similar trends are observed in each case.

### 3.2. Evaluation of coarse attitude determination using the TRIAD method

First, the error characteristics of the conventional coarse attitude determination based on the TRIAD method are evaluated. The analysis uses the onboard-converted Sun direction uS_b and magnetic field direction uB_b, together with the corresponding model values uS_i and uB_i (b: body-fixed frame, i: inertial frame). The attitude error is evaluated using the root-mean-square (RMS) value of the rotation angle between the estimated attitude Y_pred and the reference attitude Y_true.

All data sets from passes P1 to P5 are used in this evaluation. When the magnetic field direction is treated as the more reliable vector, the RMS attitude error is 7.34 deg, while the Sun-direction error is 3.41 deg. Conversely, when the Sun direction is treated as the more reliable vector, the RMS attitude error is 7.65 deg, and the magnetic field direction error is 3.39 deg. The reason why the sensor direction errors do not directly correspond to the attitude errors is that the attitude data are based on actual measurements and therefore include sensor noise, bias errors, model errors, and alignment errors.

Within the range of data used in this analysis, no significant difference is observed regardless of which sensor is prioritized, indicating comparable performance between the two cases.

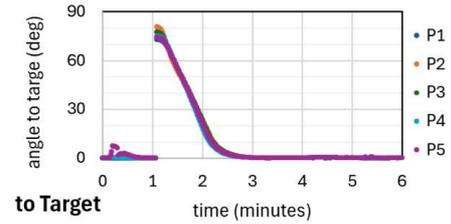

Fig. 3. Time histories of the target attitude angles (P1–P5)

### 3.3. Learning using both CSS and MAG

An error evaluation of attitude determination is performed for cases in which both CSS and MAG measurements are used for learning. In all cases, the input X used to estimate the attitude at a single time step Y consists of measurements from the previous five time steps (i.e., −4 s to 0 s).

The definitions of the evaluated cases are summarized in Table 1. Case C1a uses three CSS values (uS_c) and MAG values (uB_m). Case C1b extends the CSS input to six values (uS_c, uE_c). In cases C1c to C1e, model-derived vectors are additionally included.

Case C1f further incorporates angular rate measurements W_g obtained from the gyro sensor (FOG). These gyro measurements are implicitly included in the reference attitude Y, which is treated as ground truth, through sequential integration using the FOG with star sensor measurements as initial values. In the test set (P5), angular rate information is assumed to be known while the attitude itself is unknown, allowing the contribution of gyro measurements to attitude estimation to be evaluated.

Three different random seeds for initializing the Conv1D network parameters (R1–R3) are prepared, and cases C1a to C1f are evaluated for each seed.

The results are summarized as follows. Using models trained on passes P1 to P4 (training data), attitude determination is performed for pass P5 (test data). The rotation-angle RMS values between the estimated attitude array Y_pred and the ground truth array Y_test are reported. For each case, the minimum RMS value among the three random seeds is highlighted with yellow hatching.

In the training set, the best performance is achieved by C1e and C1f (0.7 deg), followed by C1b (0.9 deg). The inclusion of model-derived vectors in C1c to C1f does not necessarily guarantee superior performance.

The RMS values for the test set are also listed. Comparison of the highlighted values shows that the best-performing case in the training set does not always yield the best performance in the test set. The minimum RMS in the test set is obtained for C1e and C1f (3.0 deg).

When the RMS values for the training and test sets are combined, C1e and C1f show the best overall performance. For Case 1, the results indicate that the inclusion or exclusion of gyro measurements does not significantly affect performance.

The training-set error represents the optimal attitude estimation error obtained by applying the ML method to all data in an offline analysis. The test-set error represents the expected error when applying the ML method to real-time onboard attitude determination. In this study, four passes are used for training, and the fifth pass, acquired at a later time, is used for testing. If attitude determination with an RMS error of approximately 3.0 deg can be achieved based on past experience from an unknown initial attitude state, the method can be applied to phases such as guiding the spacecraft toward an optimal attitude for initial star sensor acquisition or performing maneuvers toward a ground station for high-speed communication without relying on a star sensor.

### 3.4. Learning using either CSS or MAG alone

In the classical TRIAD method, attitude angles cannot be determined unless at least two different vectors and their corresponding model values are available. In contrast, the Case 1 results indicate that, even when no model-derived vectors are provided—as in C1a and C1b—attitude angles can be estimated for new measurements if the relationship between the measured values and the attitude angles (used for training) is sufficiently learned. As a next step, the error characteristics are examined for cases in which only one type of measurement is available.

Case 2 (C2a–C2f) is defined by removing the magnetic field vectors (uB_m and uB_i) from C1*. Case 3 (C3a–C3f) is defined by removing the CSS measurements (uS_c, uE_c) and uS_i from C1*. The same random seeds (R1–R3) used in the Case 1 analysis are applied. For reference, C4f, in which attitude determination is performed using only the gyro measurement W_g, is also evaluated.

The results for C2a–C2f are summarized in Table 2. Cases that exceeded the maximum number of training epochs (240) are excluded from the minimum RMS evaluation. In the training set, the minimum RMS is obtained for C2b (1.3 deg), whereas in the test set, the minimum RMS is obtained for C2f (2.8 deg). Because C2b does not achieve good performance in the test set, it is evaluated as having low generalization capability. In all cases, C2a yields a large RMS, indicating that attitude determination using uS_c alone is disadvantageous. A new finding is that the Earth-side Sun sensor measurement (uE_c), which had been omitted in the TRIAD-based approach, contributes to attitude determination. Based on the combined RMS of the training and test sets, C2f is identified as the best case.

For C3a–C3f and C4f, the results summarized in Tables 3 and 4 are obtained. The minimum RMS in the training set is achieved by C3c (0.7 deg), while the minimum RMS in the test set is achieved by C3f (2.1 deg). In terms of the combined RMS of training and test sets, C3a shows a notably large value, exhibiting the same trend as C2a, which also relies on a single sensor type. Using two sensor measurements, or one sensor measurement combined with a model-derived vector, is recognized to reduce RMS to some extent.

The results obtained for C3c–C3f—with RMS values of 0.7–1.4 deg for the training set and 2.1–5.1 deg for the test set—are

Table 1. Definition of input data for Case 1

| Case | total ch | X includes | | | | | | |
|------|----------|------|------|------|------|------|------|------|
|      |          | uS_c | uB_m | uE_c | uS_i | uB_i | uE_i | W_g |
| C1a  | 6        | Y    | Y    | N    | N    | N    | N    | N   |
| C1b  | 9        | Y    | Y    | Y    | N    | N    | N    | N   |
| C1c  | 12       | Y    | Y    | N    | Y    | Y    | N    | N   |
| C1d  | 15       | Y    | Y    | N    | Y    | Y    | Y    | N   |
| C1e  | 18       | Y    | Y    | Y    | Y    | Y    | Y    | N   |
| C1f  | 21       | Y    | Y    | Y    | Y    | Y    | Y    | Y   |

Table 2. Summary of results for Case 1

| Case | Train Set: Att. error RMS (deg) | | | | Test Set: Att. error RMS (deg) | | | | (I)+(II) |
|------|------|------|------|------|------|------|------|------|------|
|      | R1 | R2 | R3 | MIN (I) | R1 | R2 | R3 | MIN (II) | |
| C1a | 6.9 | 1.5 | 1.9 | **1.5** | 9.7 | 5.0 | 5.2 | **5.0** | 6.6 |
| C1b | 2.0 | 0.9 | 4.4 | **0.9** | 3.7 | 4.4 | 7.5 | **3.7** | 4.6 |
| C1c | 1.1 | 1.3 | 1.1 | **1.1** | 4.1 | 3.7 | 4.3 | **3.7** | 4.8 |
| C1d | 1.1 | 1.2 | 1.1 | **1.1** | 5.4 | 4.1 | 4.6 | **4.1** | 5.2 |
| C1e | 0.7 | 2.8 | 6.5 | **0.7** | 3.0 | 3.9 | 7.3 | **3.0** | 3.7 |
| C1f | 2.2 | 0.7 | 1.8 | **0.7** | 3.8 | 3.0 | 3.5 | **3.0** | 3.7 |

Table 3. Results for Case 2 (excluding uB_m and uB_i from Case 1)

| Case | Train Set: att. error RMS (deg) | | | | Test Set: att. error RMS (deg) | | | | (I)+(II) |
|------|------|------|------|------|------|------|------|------|------|
|      | R1 | R2 | R3 | MIN (I) | R1 | R2 | R3 | MIN (II) | |
| C2a | 6.4 | 7.0 | 3.8* | **6.4** | 8.6 | 7.6 | 10.0* | **7.6** | 13.9 |
| C2b | 1.6 | 1.3 | 1.9 | **1.3** | 5.6 | 6.5 | 6.4 | **5.6** | 6.9 |
| C2c | 4.9 | 4.1 | 3.8 | **3.8** | 4.9 | 5.1 | 5.7 | **4.9** | 8.7 |
| C2d | 3.5 | 4.3 | 3.7 | **3.5** | 3.8 | 3.9 | 3.9 | **3.8** | 7.3 |
| C2e | 2.5 | 1.6 | 1.7 | **1.6** | 6.2 | 5.6 | 5.6 | **5.6** | 7.1 |
| C2f | 1.7 | 1.5 | 1.7 | **1.5** | 3.0 | 2.8 | 3.0 | **2.8** | 4.3 |

* = stopped at max epoch

Table 4. Results for Cases 3 and 4 (Cases C3* exclude uS_c, uE_c, and uS_i from Case 1; Cases C4* exclude all inputs except W_g)

| Case | Train Set: att. error RMS (deg) | | | | Test Set: att. error RMS (deg) | | | | (I)+(II) |
|------|------|------|------|------|------|------|------|------|------|
|      | R1 | R2 | R3 | MIN (I) | R1 | R2 | R3 | MIN (II) | |
| C3a | 3.5 | 4.0 | 3.6 | **3.5** | 13.9 | 14.7 | 14.6 | **13.9** | 17.4 |
| C3c | 3.2 | 0.7 | 1.4 | **0.7** | 10.8 | 5.9 | 5.1 | **5.1** | 5.8 |
| C3d | 1.4 | 2.2 | 1.8 | **1.4** | 4.4 | 4.8 | 4.2 | **4.2** | 5.6 |
| C3f | 1.0* | 1.1 | 1.1 | **1.1** | 2.8* | 2.1 | 3.6 | **2.1** | 3.2 |
| C4f | 3.8* | 5.0 | 4.1 | **4.1** | 7.6* | 10.1 | 7.1 | **7.1** | 11.2 |

* = stopped at max epoch

considered comparable to those of C1c–C1f. These results suggest the possibility that a certain level of attitude determination accuracy can be achieved using only magnetic field measurements in eclipse conditions where sunlight is unavailable.

Case C4f, which uses only gyro measurements, is evaluated as a reference and yields a relatively large combined RMS, following C3a and C2a. Although attitude determination from angular rate information alone without attitude initialization is theoretically impossible, the model is able to estimate attitude angles by exploiting similarities in numerical patterns within the training data.

Overall, among Cases C1–C4, the minimum RMS in the training set is achieved by C1e, C1f, and C3c (0.7 deg), while the minimum RMS in the test set is achieved by C3f (2.1 deg). The smallest combined RMS is also obtained for C3f (3.2 deg). When gyro measurements are excluded (excluding C*f), C1e yields the best test-set RMS (3.0 deg), and C3e yields the best combined RMS (3.2 deg). These results indicate that, when a high-precision gyro sensor is available, attitude determination using magnetic field and gyro sensors is optimal. In the absence of a gyro sensor, the combination of Sun sensors and magnetic field sensors provides favorable performance. When only magnetic field sensors are available during eclipse, the selectable options are C3a–C3d, among which C3d yields the best result (combined RMS of 5.6 deg), albeit with increased error.

### 3.5. Evaluation of time-series data

One reason why ML-based attitude determination functions effectively in this study is that the five observation passes exhibit strong similarities. In actual satellite operations, such similarity naturally arises when attitude maneuver strategies are standardized and observations over similar regions are conducted using the same operational scenarios. To illustrate the degree of similarity, excerpts of the time-series data for CSS and MAG measurements are shown in Fig. 4.

The CSS and MAG measurements for pass P1 (training set) and pass P5 (test set) are compared. For CSS, the components CSS[1], CSS[4], and CSS[2] have larger values than the other three components. Although the offsets differ, the geometric shapes of the time-series profiles are similar. For MAG, the output trends of each axis are also evaluated to be highly similar between P1 and P5. Because the time-series data are shuffled during training, long-term temporal trends are not explicitly included in the learning process.

Next, for P1 (training set) and P5 (test set), the estimated attitude (ATT) and sensor vectors (Sun, Mag, Earth) are evaluated by comparing the differences between the estimated values (pred) and the ground truth (true) as time-series data. The learning model C1e (random seed R1; RMS = 0.7 deg during training and 3.0 deg during testing) is used for this evaluation.

For the attitude error (ATT), shown by the black line, the errors for P1 (training set) are distributed within approximately 3.1 deg, whereas the maximum error for P5 (test set) reaches 8.2 deg. In the central portion of the time series, an offset of approximately 3 deg persists, while larger errors are observed near the beginning and the end of the sequence.

For the sensor vector errors, the smallest errors are observed for the Sun vector, followed by the magnetic field vector, and then the Earth vector. It is inferred that the magnetic field and Earth vectors exhibit larger nonlinearity and offsets due to factors such as variations in the internal magnetic environment of onboard equipment and the complexity of Earth albedo.

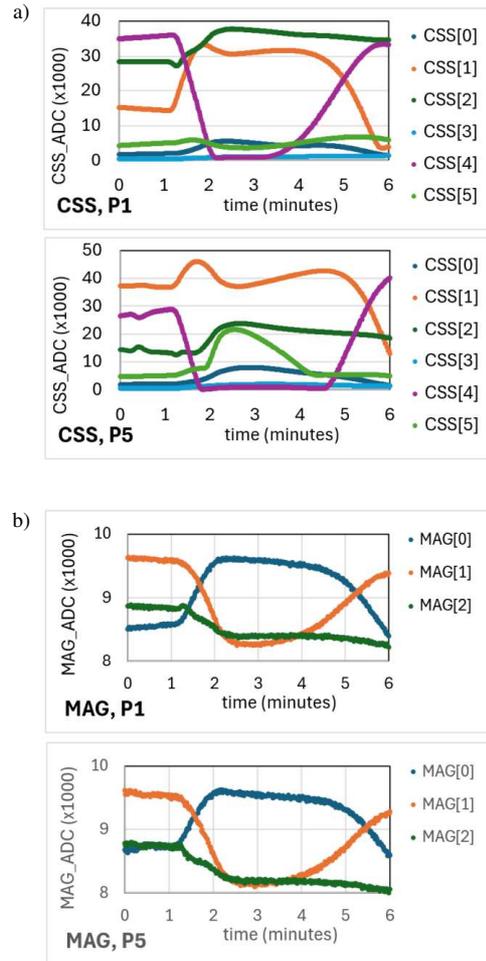

Fig. 4. Comparison of time-series CSS and MAG values (excerpts from P1 and P5)

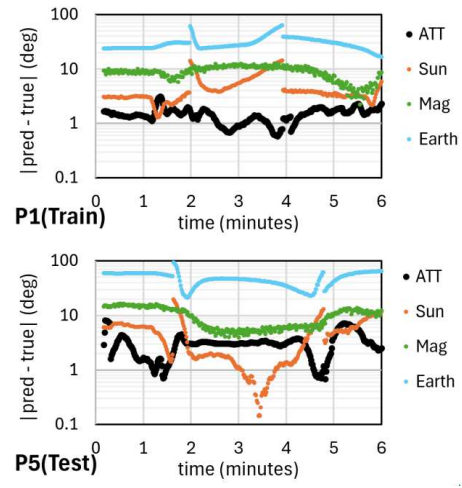

Fig. 5. Time-series evaluation of attitude estimation results (excerpts from P1 and P5)

### 4. Conclusion

In this paper, attitude determination performance was re-evaluated by applying a new analysis method to flight data obtained from a microsatellite whose mission has already been completed. A Conv1D-based approach was employed to regressively predict attitude from consecutive time-series data. Observation logs from the same region were intentionally used to align conditions and facilitate evaluation. While the RMS attitude error obtained using the TRIAD method is approximately 7 deg, the ML-based approach achieved RMS errors of 0.7 deg (training set; e.g., Case C1e) and 2.1 deg (test set; Case C3f). Although the accuracy degrades when using a single sensor alone, the inclusion of model-derived vectors improves performance. The results also indicate that a certain level of attitude determination is possible using only magnetic field measurements during eclipse conditions.

Based on these findings, ML-based methods are recognized as promising candidates for both onboard computation in future satellites and offline analysis. Because the available range of flight log data is limited, constructing learning models through simulator-based validation is considered valuable. In addition, operational strategies such as selecting among multiple models depending on mission conditions are expected to play an important role.

It should be noted that the results presented in this study are based on a supervised learning model in which high-quality attitude determination values obtained from star sensors and a fiber optical gyro (FOG) are treated as ground truth. These results do not imply that attitude determination is generally possible using only a single coarse attitude sensor. In this analysis, nonlinear sensor biases, alignment errors, and spacecraft-specific disturbance characteristics present in the actual flight environment are implicitly embedded in the learning model through the ground-truth data, enabling attitude estimation under conditions similar to those of past operational data. In contrast, in situations where no ground truth is available, even if a learning model is constructed in advance based on simulation data, reliable attitude determination using only a single coarse attitude sensor is considered difficult unless nonlinear errors specific to the real environment can be accurately identified. Therefore, the outcomes of this study demonstrate the potential applicability of the proposed approach to post-flight analysis using ground truth data, as well as to auxiliary purposes such as initial acquisition support or redundancy assistance, rather than generalizing the attitude determination problem beyond fundamental observability constraints.

The motivation for focusing on the combination of coarse attitude sensors and machine learning in this study arises from the practical limitation that, in many CubeSats, it is difficult to install high-precision attitude sensors, making it challenging to directly and continuously measure the true attitude. At the same time, the results of this study suggest that constructing a meaningful learning model requires some form of ground truth or information closely related to it. Such information does not necessarily have to be a direct attitude ground truth obtained from high-precision star sensors; even intermittently available information that provides clues for attitude estimation may function as an effective supervisory signal for learning. For example, fisheye camera images capturing the Earth direction or Earth observation images in which ground coordinates can be recognized may serve as indirect ground-truth information that imposes constraints on attitude estimation. Future work should focus on integrating such image-based information with other observational data to establish attitude estimation methods that can be trained even under conditions where direct attitude ground truth is unavailable.

**bibliographic note**